# Bubble Shape Oscillations and the Onset of Sonoluminescence

Michael P. Brenner [†], Detlef Lohse [*] and T.F. Dupont

*Department of Mathematics, The University of Chicago*

*5734 University Ave., Chicago, IL 60637, USA*

(June 27, 1995)

An air bubble trapped in water by an oscillating acoustic field undergoes either spherical or nonspherical pulsations depending on the strength of the forcing pressure. Two different instability mechanisms (the Rayleigh–Taylor instability and parametric instability) cause deviations from sphericity. Distinguishing these mechanisms allows explanation of many features of recent experiments on single bubble sonoluminescence, and suggests methods for finding sonoluminescence in different parameter regimes.

The discovery and elucidation of single bubble sonoluminescence (SL) [1–3] sparked a renewal of interest in the dynamics of a levitated bubble driven by an oscillating acoustic field. Detailed optical measurements [4,5] (cf. figure 2 of [4]) reveal different dynamic regimes, distinguished by the bubble shape, which depend on the ambient bubble radius $R_0$ and the strength of the forcing pressure amplitude $P_a$. For very small forcing, the bubble is spherical throughout the oscillation period. At higher forcing the bubble develops non–spherical surface oscillations. Two different instability mechanisms are operating: (i) The Rayleigh–Taylor (RT) instability [6], occurring whenever gas is strongly accelerated into a liquid. (ii) Parametric instability, arising due to the accumulation of perturbations from sphericity over many oscillation periods.

However, experiments show several phenomona which are confusing in light of these two instabilities: (a) Near the onset of sonoluminescence for air bubbles in water, the nonspherical pulsations vanish [1,4,5] and the bubble regains spherical symmetry. This is at variance with the RT instability, which should be more potent at the stronger accelerations required for SL. (b) In highly viscous fluids, the parametric instability is diminished. Nevertheless, it has proven extremely difficult (if not impossible) to make bubbles in highly viscous fluids (greater than about ten times the viscosity of water) [1] undergo strong oscillations, even at very strong forcing.

The purpose of this paper is to assist in resolving these puzzles by presenting a stability analysis [7,8] of the oscillating bubble. Our main result is that we account for both above experimental phenomena (a) and (b) by taking into consideration the detailed interplay between the two instabilities (i) and (ii). All parameters correspond to the experimental ones [3,4,9] for an air bubble in water: the surface tension of the air-water interface is $\sigma = 0.073 kg/s^2$, the viscosity of water is $\nu = 10^{-2} cm^2/s$, its density is $\rho_w = 1000 kg/m^3$, the speed of sound in water is $c_w = 1481 m/s$, the driving frequency of the forcing acoustic field $P(t) = P_a \cos(\omega t)$ is $\omega/2\pi = 26.4 kHz$, the external pressure $P_0 = 1 atm$, and finally $\gamma = 1.4$ for the ratio of the specific heats.

The radius $R(t)$ of a driven spherical bubble obeys the Rayleigh–Plesset (RP) equation [10,9,8]

$$R\ddot{R} + \frac{3}{2}\dot{R}^2 = \frac{1}{\rho_w}\left(p(R,t) - P(t) - P_0\right) + \frac{R}{\rho_w c_w}\frac{d}{dt}\left(p(R,t) - P(t)\right) - 4\nu\frac{\dot{R}}{R} - \frac{2\sigma}{\rho_w R}. \quad (1)$$

Löfstedt, Barber, and Putterman show that if $p(R,t)$ is adiabatically slaved to the bubble radius via a van der Waals equation of state, i.e.

$$p(R) = P_0 \frac{\left(R_0^3 - h^3\right)^\gamma}{\left(R^3 - h^3\right)^\gamma}, \quad (2)$$

then the Rayleigh–Plesset equation (1) well describes the experimental $R(t)$ [9]. Here, $h = R_0/8.54$ (for air) is the hard core van der Waals radius. The adiabatic RP equation applies whenever $\dot{R} < c_{gas}$, where $c_{gas}$ is the speed of sound in the gas [9].

We now focus on the stability of the radial solution $R(t)$. Following previous authors [7,11,8], consider a small distortion of the spherical interface $R(t) + a_n(t)Y_n$ where $Y_n$ is a spherical harmonic of degree $n$. The dynamics for the distortion amplitude $a_n(t)$ is given by

$$\ddot{a}_n + B_n(t)\dot{a}_n - A_n(t)a_n = 0. \quad (3)$$

The full expressions for $A_n(t)$ and $B_n(t)$ for a gas bubble in a viscous fluid were first derived by Prosperetti [8]; they are nonlocal in time, reflecting the interaction between the bubble and the flow in the fluid, initiated by the bubble wall motion itself. Our calculations employ a local approximation of the full hydrodynamic equations (17), (23), and (25) of [8], assuming that there will be fluid flow only in a thin layer of thickness $\delta$ around the bubble. Approximate solutions of the full hydrodynamic equations [12] give the penetration depth $\delta$ as the minimum of the diffusion length scale $\sqrt{\nu/\omega}$ and of $R/(2n)$. Then

$$A_n(t) = (n-1)\frac{\ddot{R}}{R} - \frac{\beta_n \sigma}{\rho_w R^3}$$

$$-\left((n-1)(n+2)+2n(n+2)(n-1)\frac{\delta}{R}\right)\frac{2\nu\dot{R}}{R^3}, \quad (4)$$

and

$$B_n(t) = 3\frac{\dot{R}}{R} + \left((n+2)(2n+1) - 2n(n+2)^2\frac{\delta}{R}\right)\frac{2\nu}{R^2}, \quad (5)$$

where $\beta_n = (n-1)(n+1)(n+2)$. Note that the second term in eq. (5) causes damping [13] of the shape oscillation and is always positive. The penetration depth takes into account both radial and azimuthal diffusive fluxes in the velocity field. The azimuthal gradients ensure that the viscous contribution to $B_n(t)$ stabilize the bubble.

*Rayleigh–Taylor Instability:* First we focus on the RT instability, which occurs near the minimum bubble radius where the acceleration $\ddot{R}$ of the bubble wall is positive (i.e., the gas accelerates into the fluid). Nonspherical perturbations of the bubble shape grow during this time period. The amplification factor follows from a WKB type analysis on equation (3), taking $a_n(t) \sim \exp(S_n(t))$. The average amplification can be estimated as

$$\bar{S}_n \approx \int \dot{S}_n(t) dt = \int \left(-\frac{B_n}{2} \pm \sqrt{\frac{B_n^2}{4} + A_n}\right) dt \quad (6)$$

where the integral is evaluated over the time period during which $\dot{S}_n(t)$ is positive. Appreciable growth occurs when molecular fluctuation (size 1nm) can grow to the minimum size of the bubble. Figure 1 shows the phase diagram of this instability as a function of the ambient bubble radius $R_0$ and the forcing pressure $P_a$.

*Parametric Instability:* After transient effects, $R(t)$ and thus $A_n(t)$ and $B_n(t)$ are periodic with period $T = 2\pi/\omega$ [14]. Eq. (3) is then called Hill's equation. Parametric instability corresponds to a net growth of a nonspherical perturbations each oscillation period, so that after many periods perturbations overwhelm the bubble. Formally, this occurs whenever the magnitude of the maximal eigenvalue of the Floquet transition matrix $F_n(T)$ is larger than one. $F_n(T)$ is defined by

$$\begin{pmatrix} a_n(T) \\ \dot{a}_n(T) \end{pmatrix} = F_n(T) \begin{pmatrix} a_n(0) \\ \dot{a}_n(0) \end{pmatrix}. \quad (7)$$

By numerically computing $F_n(T)$ and determining its eigenvalues, we mapped out the entire phase diagram of stability. We first calculate the stability diagram for the $n=2$ mode at zero viscosity (fig. 2). In general, when $P_a$ and $R_0$ are large, the bubble becomes more unstable, although the detailed structure of the phase diagram is quite complicated. Many features can be understood analytically [11] by examining the small forcing ($P_a \ll \gamma P_0$) limit, where eq. (3) can be reduced to Mathieu's equation. Substituting $b_n(t) \propto (R(t))^{3/2} a_n(t)$ one obtains

$$b_n'' + 2\xi_M b_n' + \omega_M^2 (1 + \epsilon_M \cos(\tilde{t})) b_n = 0, \quad (8)$$

with $\omega_M^2 = \beta_n \sigma/(\rho_w R_0^3 \omega^2)$, $\epsilon_M = P_a/(\gamma P_0)$ and $\xi_M = [(n+2)(2n+1) - 2n(n+2)^2 \delta/R_0]\nu/(R_0^2 \omega)$. Primes denote derivation with respect to the dimensionless time $\tilde{t} = \omega t$. The well known Mathieu tongues [15] occur in fig. 2 at $\omega_M = k/2$, where $k$ is an integer. Since $\omega_M \propto R_0^{-3/2}$ the tongues are not equidistant but become more and more packed at small $R_0$. For $n=2$ the first Mathieu instability ($k=1$) occurs at $R_0 = 50\mu m$.

Finite viscosity $\xi_M \propto \nu \neq 0$ stabilizes the surface dynamics. Though $\xi_M$ is small ($\xi_M \sim 10^{-4}$ for $R_0 = 10\mu m$), the Mathieu tongues are stabilized [16]. The stabilization is stronger at small $R_0$, where $\omega_M$ and $\xi_M$ are large. Figure 1 shows a superposition of the stability diagrams for modes $n = 2, 3, 4, 5, 6$ corresponding to the viscosity of water $\nu = 10^{-2} cm^2/s$. Stability diagrams of this type were first considered by Eller and Crum [11], and later by Horsburgh and Holt [17]. These studies examine larger bubble sizes than shown in fig. 1; our calculations in the large bubble regime [12] give similar thresholds as found in [11,17]. A particularly interesting new feature of our fig. 1 is the presence of small islands of stability dispersed throughout the unstable domain.

When the bubble oscillations are weak, there is saturation of the linear instability, leading to an oscillating nonspherical bubble. Such shapes have been observed [11,17]. In the strongly nonlinear regime, saturation is unlikely, since parametric instability causes $a_n(t)$ of roughly constant amplitude throughout a cycle, whereas the bubble size $R(t)$ changes by orders of magnitude. The RT instability causes $10^4$ growth of perturbations on timescales of less than $10^{-9}$s, which almost certainly leads to the destruction of the bubble.

What are the consequences of these results for the SL experiment? A standard experimental protocol [18,1,4] is to slowly increase the driving pressure $P_a$ for a bubble of given ambient radius $R_0$. For low viscosity fluids (c.f. fig. 1) the parametric instability sets in *before* RT. The parametric instability acts over a long time scale ($\sim 10$ cycles) which is comparable to the diffusive timescale. The bubble therefore has time to readjust its size $R_0$ by enhancing diffusion through the nonspherical surface, thereby reentering a stable parameter region [19]. In this way the long ($10^{-3}$s) time scale parametric instability "protects" the bubble from encountering a region of the phase diagram where it could be destroyed by the ($10^{-9}$s) RT instability. Upon further increasing $P_a$, the same mechanism works again: The bubble more or less tracks the parametric instability borderline in the phase diagram until finally the parameter regime is reached where the forcing is strong enough so that a shock can be emitted from the collapsing bubble wall. The shock is associated with the sonoluminescence [21] and experimentally starts at $P_a \approx 1.15$atm [4]. In the strong forcing regime there may be additional mechanisms associated with the sonoluminescence itself that help stabilize the bubble [22].

An immediate consequence of above scenario is that the bubble dynamics should be *hysteretic*: On increasing the forcing pressure, the bubble undergoes nonspherical oscillations. At an even slightly higher forcing pressure, the bubble will shrink. If the forcing pressure is then slowly decreased, the oscillations will become spherical again, even beyond that pressure where nonsphericity initially appeared. This type of hysteretic behavior was noted in the experiments of Gaitan [1].

We next discuss how increasing the viscosity changes the above scenario. A large viscosity increases the parametric instability thresholds: at a viscosity of ten times that of water, the entire parameter range shown in figure (1) is stable. On the other hand, the Rayleigh–Taylor instability depends only weakly on viscous effects, because the maximum acceleration of the bubble is essentially independent of viscosity [9]. Our numerical calculations show the RT threshold changes only slightly. Thus in highly viscous fluids (as glycerol), the bubble encounters the RT instability at a *smaller* driving pressure than the parametric instability, rather than the other way round, as in water. The RT instability acts so quickly ($10^4$ amplification within $10^{-9}s$) that the bubble has no time to reenter a stable part of parameter space before being completely destroyed. We hypothesize that this could explain why SL has not yet been observed in highly viscous fluids. However, note that this problem could be surmounted by preparing or choosing bubbles of small enough ambient radius that as the driving pressure is continuously increased, the bubble does not experience the Rayleigh Taylor instability before sonoluminescing. An alternative explanation for the absence SL in glycerol relies on the very different solubility of air in water and glycerol, respectively [20]; we cannot rule this out.

We again point out that our calculations depend on the basic assumption that the pressure is adiabaticity slaved to the bubble radius. This only holds when the bubble wall velocity is subsonic. For an air bubble in the $R_0$ range of interest, adiabatic calculations show that the bubble wall becomes supersonic at $P_a \sim 1.35$atm. The experiments [4,23] show that sonoluminescence begins already around $P_a \sim 1.15$atm (indicating the onset of some shock [21]). We take this as a hint that a more complete treatment should take heat diffusion [24,9] into account, since decreasing $\gamma$ decreases the speed of sound in the gas. Work is in progress [12].

In conclusion, we have analyzed the stability of a cavitating bubble to both Rayleigh Taylor and parametric instabilities. At the high forcing pressures necessary for SL, only a bubble with a very small ambient radius is stable to both the RT and parametric instability. For low viscosity fluids, the parametric instability "pushes" the bubble to this special region by forcing the ambient size of the bubble to decrease. For more viscous fluids the parametric instability does not exist in the appropriate parameter region, so the bubble is destroyed by the RT instability before sonoluminescing. At the heart of the argument are the very different timescales on which the RT ($10^{-9}$s) and the parametric instability ($10^{-3}$s) act. Finally, our scenario also accounts for hystersis.

Many mysteries remain: Are there additional mechanisms for stabilizing the bubble wall which arise from the sonoluminescence itself? What is the nature of the ambient radius dynamics at a fixed forcing pressure? What is the role of mass transport mechanisms beyond diffusion [20]? Why are bubbles with inert gases so much stabler than nitrogen bubbles?

**Acknowledgements:** We are grateful to R. Almgren, S. Grossmann, R. Holt, L. Kramer, W. Lauterborn, S. Putterman and his group, and H. J. Stöckmann for valuable hints and to L.P. Kadanoff for continued encouragement. This work has been supported by DOE contract number DE-FG02-92ER25119 and by the MRSEC Program of the National Science Foundation under Award Number DMR-9400379.
† e-mail: brenner@cs.uchicago.edu. Address after August 15, 1995: Dept. of Mathematics, Massachusetts Institute of Technology, Cambridge, MA 02139.
∗ e-mail: lohse@cs.uchicago.edu; on leave of absence from Fachbereich Physik, Universität Marburg, Renthof 6, D-35032 Marburg.

FIG. 1. Phase diagram for finite viscosity $\nu = 10^{-2} cm^2/s$. $R_0$ is given in $\mu$m and $P_a$ in atm. The black region indicates stability with respect to all modes $n = 2, 3, 4, 5, 6$. The diagram is strongly dominated by the instability of the $n = 2$-mode and considering even higher modes ($n > 6$) does not change the figure. Also shown is the RT stability curve (solid line) beyond which the bubble is RT unstable. Parameters are given in the text.

FIG. 2. The phase diagram of steady state bubble motions for $n = 2$ and $\nu = 0$. All other parameters are as in fig. 1. The black region indicates stability, and the white region instability.

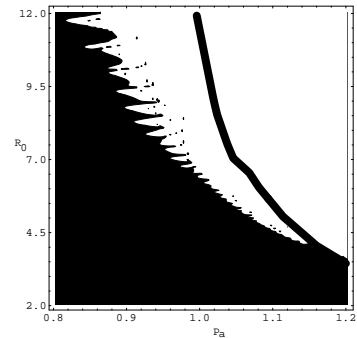

- Fig. 1

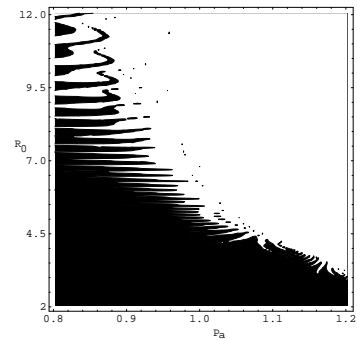

- Fig. 2